\documentclass[journal]{vgtc}         

\usepackage{mathptmx}
\usepackage{graphicx}
\usepackage{times}

\usepackage{subfigure}

\usepackage[bookmarks,backref=true,linkcolor=black]{hyperref} 
\hypersetup{
  pdfauthor = {},
  pdftitle = {},
  pdfsubject = {},
  pdfkeywords = {},
  colorlinks=true,
  linkcolor= black,
  citecolor= black,
  pageanchor=true,
  urlcolor = black,
  plainpages = false,
  linktocpage
}

\onlineid{0}

\newcommand{\remove}[1]{}

\vgtccategory{Research}

\vgtcinsertpkg

\title{Adding Semantic Information into Data Models by Learning Domain Expertise from User Interaction}

\author{Nathan O. Hodas, Alex Endert}
\authorfooter{
\item
Nathan Hodas is with Pacific Northwest National Laboratory. E-mail: nathan.hodas@pnnl.gov.
\item
Alex Endert is with Georgia Tech. E-mail: endert@gatech.edu.

}

\shortauthortitle{Hodas, Endert: Adding Semantic Information}

\abstract{
Interactive visual analytic systems enable users to discover insights from complex data. Users can express and test hypotheses via user interaction, leveraging their domain expertise and prior knowledge to guide and steer the analytic models in the system. \remove{Model steering is a more commonly used approach to integrating this human feedback into the system, creating joint cognitive systems.} For example, semantic interaction techniques enable systems to learn from the user's interactions and steer the underlying analytic models based on the user's analytical reasoning. However, an open challenge is how to not only steer models based on the dimensions or features of the data, but how to add dimensions or attributes to the data based on the domain expertise of the user. In this paper, we present a technique for inferring and appending dimensions onto the dataset based on the prior expertise of the user expressed via user interactions. Our technique enables users to directly manipulate a spatial organization of data, from which both the dimensions of the data are weighted, and also dimensions created to represent the prior knowledge the user brings to the system. We describe this technique and demonstrate its utility via a use case. 
} 


\CCScatlist{ 
 \CCScat{K.6.1}{Management of Computing and Information Systems}%
{Project and People Management}{Life Cycle};
 \CCScat{K.7.m}{The Computing Profession}{Miscellaneous}{Ethics}
}

\begin{document}


\firstsection{Introduction}

\maketitle

Visual analytic systems facilitate a cognitive process of discovery by combining data analytic and information visualization capabilities \remove{\cite{thomas_illuminating_2005}}. These technologies leverage computational models to generate visual representations of data. Through user interaction, analysts make use of analytical reasoning to pose questions, form hypotheses, and gain insights into the data. For exploratory data analysis, this process can be referred to as sensemaking \cite{pirolli_sensemaking_2005}.

The analytic models used in visual analytics typically approximate some characteristics, metrics, attributes, or other features about the data. Thus, user interactions are typically designed to augment the analytic or visual parameters. More recently, semantic interaction techniques enable analytic model steering via inferring data characteristics that are of interest from the user based on their exploratory user interactions \cite{endert_beyond_2013}. However, one limitation of such approaches is that they attempt to fit the inference of the user's feedback into the existing feature space in the data model (e.g., keywords for text corpora). \remove{Prior work has shown that sensemaking and analytical reasoning of text corpora often entails users not only reasoning about the existing keywords in a corpora, but incorporating additional features and domain expertise into the dataset as part of their reasoning process \cite{endert_semantics_2012}\cite{andrews_space_2011}\cite{klein_making_2006}.} Analysts create mental constructs and concepts about the data which do not necessarily reflect in the data features, but extend into the broader context and features existent in the domain of the data being explored. \remove{This is important, as it is because of this ability for humans to impart their expertise that visual analytics strives for user feedback and the "human-in-the-loop" approach to data science. }
Thus, \textit{how can visual analytic systems infer such additional, semantic dimensions about the data from the user?}

In this paper, we demonstrate a system for allowing a user to create additional semantic features and leverage these features for image classification and accelerating the sense-making process.  The system presents a canvas filled with randomly arranged images to the user, and the user positions images according to their internal mental model, leveraging their previous knowledge and task-specific motivations. We call this system ActiveCanvas, because it does not simply allow the user to shuffle images around; it actively participates in the sensemaking loop by using information theoretic tools to extract relevant features from the images and then repositioning the images in an attempt to  better reflect the user's mental model. The information theoretic tools implicitly capture the analytic reasoning of the user, and it allows the system to build knowledge over time by recording the position of users and saving them as new features to be utilized in the future.

\remove{The contributions of this work include:
\begin{itemize}
\item A visual analytic technique for inferring and creating additional dimensions for datasets pertaining to the domain expertise of the user.
\item A technique for leveraging  knowledge of prior users in the visual analytic process.
\end{itemize}
}

\section{Related Work}

Spatial workspaces are effective visual metaphors for supporting sensemaking and discovery tasks. For example, Andrews et al. showed how the spatial positioning, grouping, and re-grouping of documents was an effective method to help analysts conduct a sensemaking task \cite{andrews_space_2010}. Their work found that users were able to leverage the flexibility of a large, high-resolution to organize their information. Throughout the process, the analysts made use of spatial constructs including lists, piles, sequences, and categorizations of the information. Over time, these spatial constructs evolved as the analysts gained more insights about the data. This process of incrementally refining a spatial layout based on one's own increased understanding of the data is called \textit{incremental formalism} \cite{shipman_formality_1999}. 


User interaction is critical to the success of sensemaking and discovery tasks using visual analytics. Prior work has analyzed user interaction as a data source to understand more about the user, the domain, and the data being analyzed. For example, Dou et al. showed how the analysis of user interaction logs can uncover low-level strategies and tasks of analysts performing financial analysis tasks \cite{dou_recovering_2009}. Similarly, Brown et al. analyzed user interactions for a simple search task (Where's Waldo) to discover patterns of interactions that could accurately predict fast and slow task completion times \cite{brown_finding_2014}.

\begin{figure}[h]
\centering
\begin{subfigure}[Initial Arrangement\label{fig:step1}]{\includegraphics[width=\columnwidth]{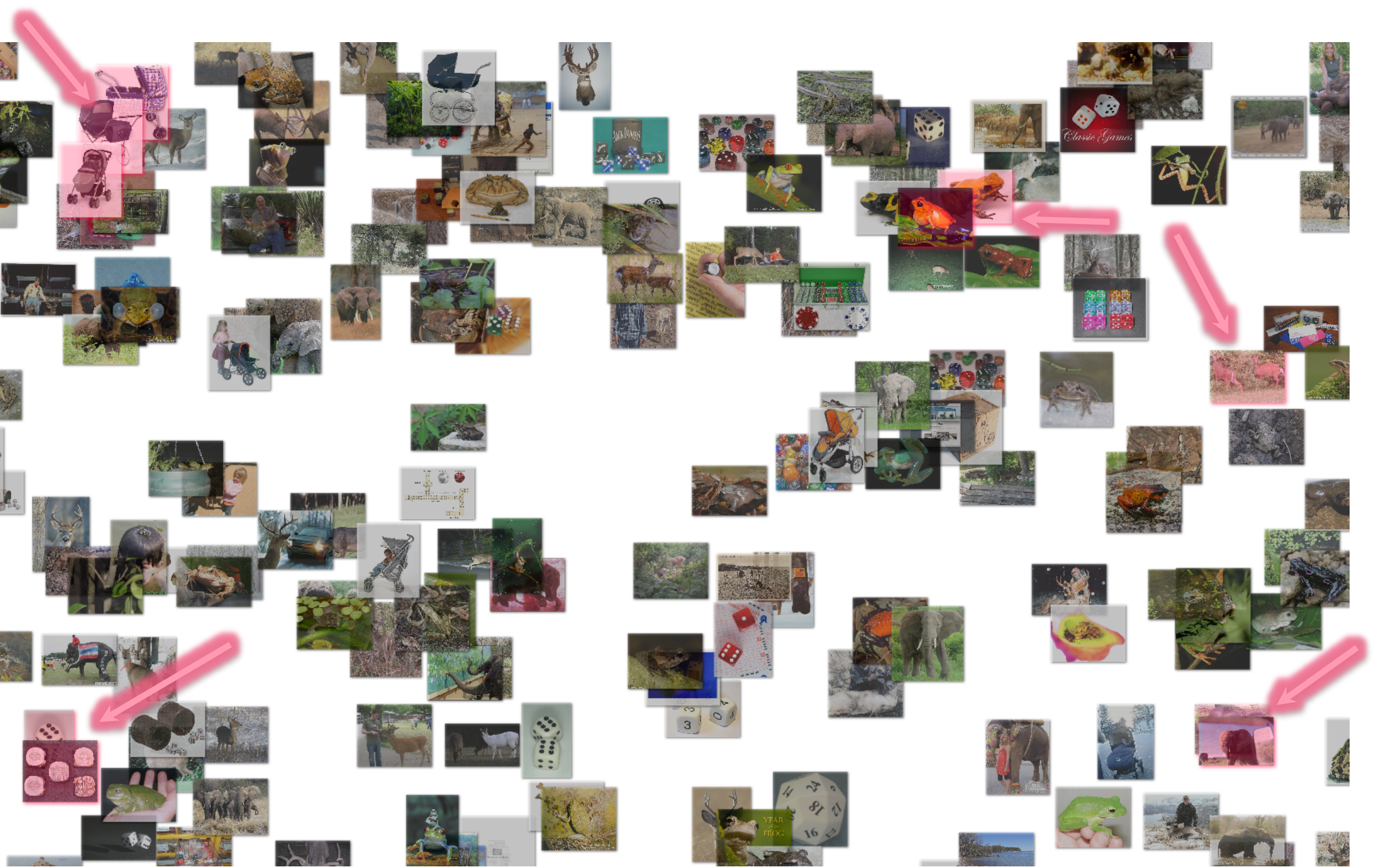}}
\end{subfigure}
\begin{subfigure}[First Refinement\label{fig:step2}]{\includegraphics[width=\columnwidth]{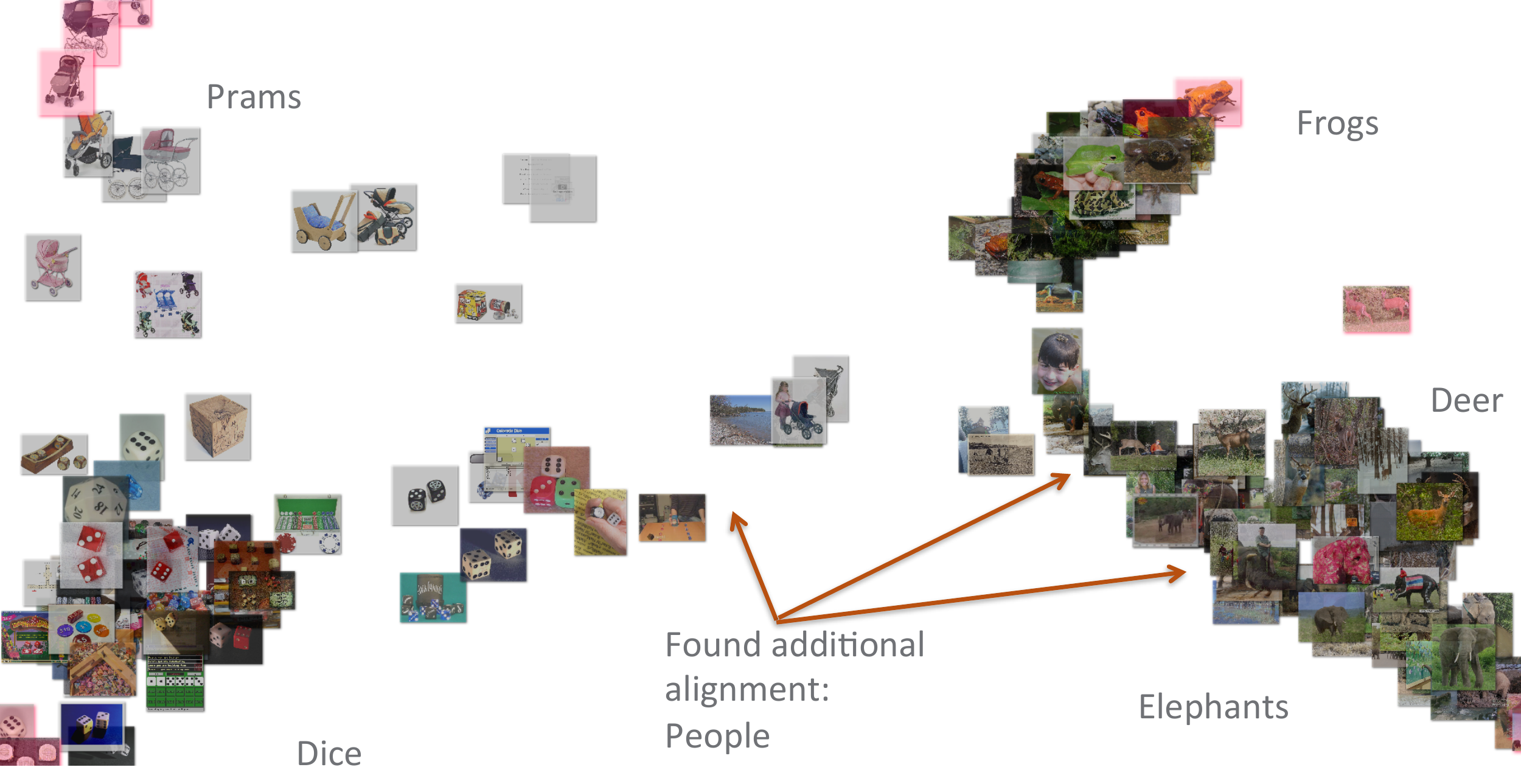}}
\end{subfigure}
\begin{subfigure}[Final Refinement, 20 images touched\label{fig:finalstep}]{\includegraphics[width=\columnwidth]{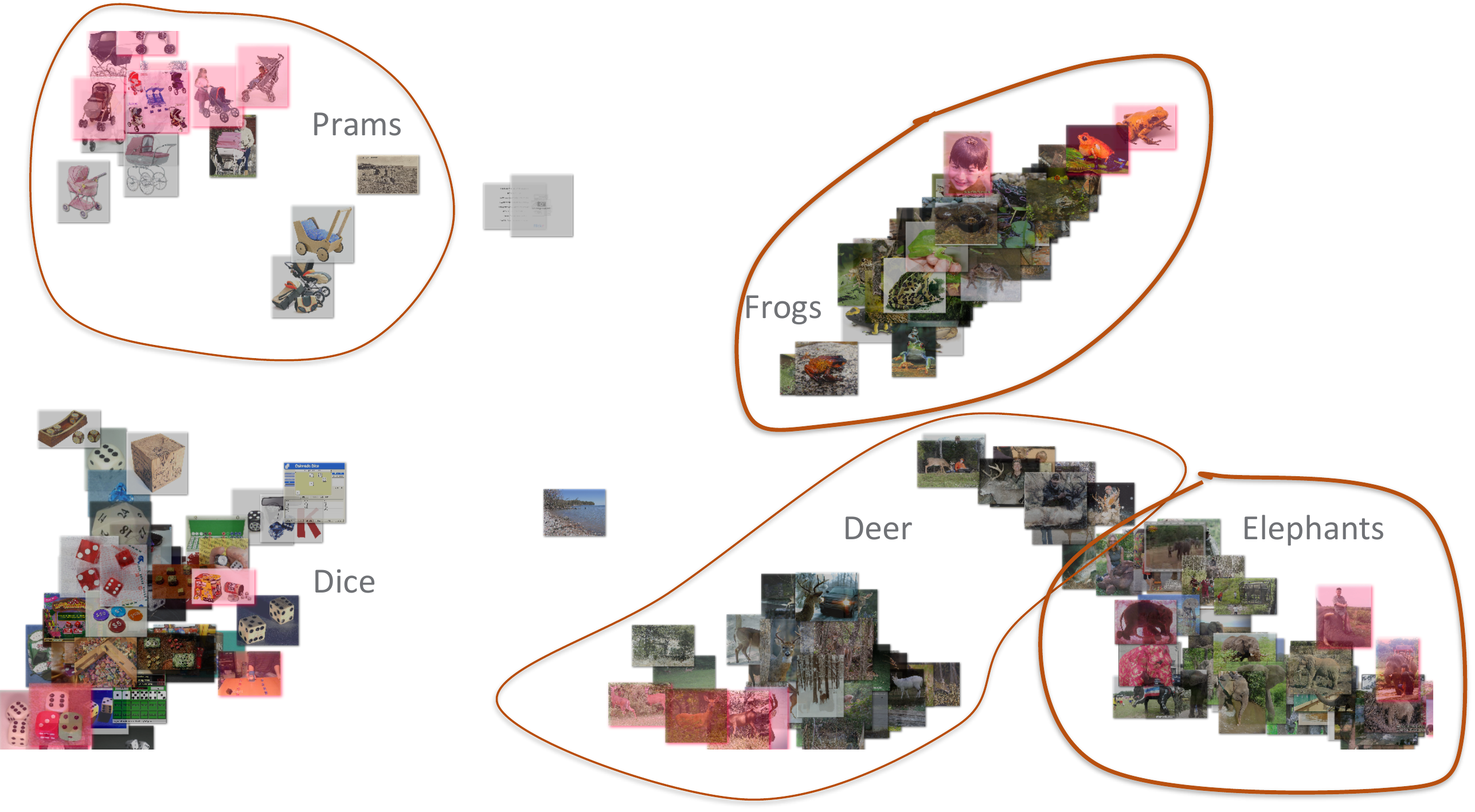}}
\end{subfigure}
\caption{An example workflow.  (a) The user initially positions 8 images (pink arrows). (b) After the first refinement, ActiveCanvas creates initial clusters. The user positions additional images to provide additional feedback to the canvas. (c) After touching a total of 20 images, and 3  refinements, clear clusters emerge. Although not readily apparent, the three images outside of clusters do not fit in any category.}\label{fig:workflow}
\end{figure}

\section{ActiveCanvas}\label{sec:discussion}

The ActiveCanvas is designed to enable the user to leverage the hard work of other users to accelerate the work of a new user sorting images on a 2d virtual space.  Imagine that a user has a latent mental model of the data they are trying to organize, denoted by $X_L$. $X_L$ represents the implicit features a user internally leverages to understand the images in front of them. Initially, the system would present them with naive arrangements of images, each of which the user would arrange on the screen to create positions $Y$.  Thus, for this limited set, $Y$ explains $X_L$, to the best of the user's intentions. \remove{Information theoretic measures are promising objective functions to evaluate this explanatory power, although they are by no means the only tool at our disposal. Information theoretic methods are often invariant to invertible transformations, so the user doesn't need to generate a model for converting $X_L$ into $Y$ nor does the system need to know this mapping to optimize mutual information.} A key assumption is that the user places items close together on the screen ($Y$-space) because they are similar in the latent space $X_L$.  Thus, if we can find a way to slightly alter the arrangement of items on the screen to maximize mutual information with the feature vector provided by the system, these new positions will likely have high mutual information with the user's mental model.  \remove{That is, the new arrangement of items on the screen explains the user's mental model, and the user's mental model explains the arrangement on the screen.}

To this end, when a user has arranged some initial images to their satisfaction, the ActiveCanvas attempts to slightly alter the new positions of those images, $Y^\prime$ such that $MI(X:Y^\prime)$ increases, where $MI$ is the mutual information. We then predict the position of the remaining items using the new positions of the touched items as a training set. using RBF SVM.  If the user is not satisfied, the user continues to adjust images or move additional images to produce better estimates of mutual information and more training data for the SVM. After the user has become satisfied with the position of all items on the screen, they may then commit the positions as two new features, changing $X \rightarrow [X Y_{final}]$. This enables other users to leverage the hard work of the first uers.

This robustness to imprecision allows a lazy user to benefit from the diligent work of more dedicated users.  If two users have classified images according to two different strategies, the lazy user need not position their images with as much rigor as would otherwise be required to distinguish clear clusters.  Upon sending the system the lazily-positioned images, ActiveCanvas leverages the features (e.g., positions chosen and recorded by the other users) to infer the intended clusters of the lazy user.  \remove{Figure~\ref{fig:goodusers} shows an example, where User 1 positions images according to the labels \{deer,dice,elephant,frog\}. User 2 positions images according to the labels \{animal,not animal\}.} After being satisfied, each user commits their arrangement back to the system, creating a new features set comprised of the original feature vector plus the (x,y) positions from each user. 

\remove{Lazy user L1 (Fig.~\ref{fig:L1}) attempts to follow the same strategy as User 1, but does not adequately separate the clusters to create distinct dividing surfaces between clusters.  Similarly, lazy user L2 attempts to follow the same strategy as User 2, but instead of following the same distinct clustering as User 2, L2 chooses a different clustering topology and roughly places animals on the inside and inanimate objects in an outside ring  surrounding the animals (but, again without creating clear separation between the two). Note that L2 has chosen to (sloppily) arrange their canvas in a different topology; User 2 made two distinct clusters, but L2 chose to make a bullseye pattern.  The mutual information measure enables ActiveCanvas to identify the user's mental model, without new users understanding or known previous user's arrangements.

The system compares L1's positions to the combined features generated by User 1 and User 2.  Based on the innate image features and similarity to the features generated by User 1, the system correctly identified the desired mental model of Lazy user L1, shown in Fig.~\ref{fig:L1final}, as well as the mental model of Lazy user L2, shown in Fig.~\ref{fig:L2final}. Thus, the value of ActiveCanvas lies in its ability to capture a user's mental model from minimal interaction from the user.
}

To exemplify this, consider the following sample workflow, worked through in Fig.~\ref{fig:workflow}.  As above, the user approaches ActiveCanvas to make sense of images chosen because they contain one of \{frog,dice,deer,elephant,pram\}, although they may be interested in the images for any reason.  Previous users have utilized ActiveCanvas on the same dataset to make clusters based on their personal desired sorting, including \{human, not human\}, \{light background, dark background\}, \{outdoors, not outdoors\}. \{animal, not animal\}, as well as according to the base classes.  After sorting the images according to their preference, they each save the positions of the images as new features. The new user approaches the canvas initially seeing a random arrangement of pictures.  They then move only 8 images into rough piles, or even singleton piles, shown in Fig.~\ref{fig:step1}.  The user then asks ActiveCanvas to refine the arrangement according to the implicit mental model ActiveCanvas uncovers from the 8 images the user touched. This results in an initial arrangement shown in Fig.~\ref{fig:step2}, where the system has separated out a number of clusters, but it has not yet reached the user's mental model.  The user moves a few more images each time, iterating through the refinement process.  After touching 20 images and asking for a total of 3 refinements, the user will see the clear clusters they were hoping to uncover, shown in Fig.~\ref{fig:finalstep}. The total elapsed time, from the first touch to the final refinement, was approximately 90 seconds.

\section{Conclusion}

Most feature generation techniques, particularly for images, do not produce semantically relevant features.  That is, features tend to be abstract numbers only meaningful in (often nonlinear) combination.  Because users tend to see images based on semantic components (e.g., indoors, flowers, daytime) and not quantitative vectors (e.g., average contrast, SIFT, gradient histograms), their spatial arrangement of images will reflect the  relationships the user identified from their domain knowledge. We present a tool to enable the user to produce those semantic arrangements and implicitly capture these data models as features for future image classification.

Future work will explore how many user-provided features essentially `cover' the semantic space.  Although there are potentially an infinite number of ways to arrange images on a canvas, users may only be able to distinguish a limited set of distinct arrangement strategies. How many different users would need to provide their own semantic features before new users would simply be referencing a superposition of existing features? If one relies on machine-generated features to cover most of the geometric or color features, would the number of semantic features needed be closer to 10 or 1000? This number is particularly relevant for evaluating the crowd-sourcing potential for this type of system.

\acknowledgments{
The authors wish to thank Nathan Baker, Landon Sego and Michael Henry. This work is part of the Signature Discovery Initiative at Pacific Northwest National Laboratory.  It was conducted under the Laboratory Directed Research and Development Program at PNNL, a multiprogram national laboratory operated by Battelle for the U.S. Department of Energy.}

\bibliographystyle{abbrv}
\bibliography{AllRefs}

\end{document}